# The microcanonical thermodynamics of finite systems: The microscopic origin of condensation and phase separations; and the conditions for heat flow from lower to higher temperatures.

D. H. E. Gross[†]
Hahn-Meitner Institute and Freie Universität Berlin,
Fachbereich Physik, Glienickerstr. 100, 14109 Berlin, Germany.
Gross@hmi.de; http://www.hmi.de/people/gross

J. F. Kenney
Russian Academy of Sciences - Joint Institute of the Physics of the Earth;
Gas Resources Corporation, 11811 N. Freeway, fl. 5, Houston, TX 77060, U.S.A.
JFK@alum.MIT.edu

Abstract:

Microcanonical thermodynamics [D. H. E. Gross, Microcanonical Thermodynamic Ensembles: Phase Transitions in "Small" Systems. (World Scientific, Singapore, 2001)] allows the application of statistical mechanics both to finite and even small systems and also to the largest, self-gravitating ones. However, one must reconsider the fundamental principles of statistical mechanics especially its key quantity, entropy. Whereas in conventional thermostatistics, the homogeneity and extensivity of the system and the concavity of its entropy are central conditions, these fail for the systems considered here. For example, at phase separation, the entropy, $S(E)$, is necessarily convex to make $e^{S(E)-E/T}$ bimodal in $E$. Particularly, as inhomogeneities and surface effects cannot be scaled away, one must be careful with the standard arguments of splitting a system into two subsystems, or bringing two systems into thermal contact with energy or particle exchange. Not only the volume part of the entropy must be considered; the addition of any other external constraint, [A. Wehrl, Rev. Mod. Phys. **50**, 221 (1978)], such as a dividing surface, or the enforcement of gradients of the energy or particle profile, reduce the entropy. As will be shown here, when removing such constraints in regions of a negative heat capacity, the system may even relax under a flow of heat (energy) against a temperature slope. Thus the Clausius formulation of the second law: "Heat always flows from hot to cold," can be violated. Temperature is not a necessary or fundamental control parameter of thermostatistics. However, the second law is still satisfied and the total Boltzmann entropy increases. In the final sections of this paper, the general microscopic mechanism leading to condensation and to the convexity of the microcanonical entropy at phase separation is sketched. Also the microscopic conditions for the existence (or non-existence) of a critical end-point of the phase-separation are discussed. This is explained for the liquid-gas and the solid-liquid transition.

---

† Author for communications.

**1.** **The definition of entropy.**

The internal energy, *U*, and the entropy, *S*, are the fundamental entities of thermodynamics. Therefore, proper understanding of them is essential.

Internal energy is defined through its changes, by the first law of thermodynamics, as the balancing entity between heat absorbed and external work performed by the system. Such definition is both intuitively clear and also consistent with a mechanistic perspective of the physical universe, and states the conservation of energy.

Entropy is defined through the two, crucially different, mechanisms of its changes: the heat exchanged externally by the system, inversely-weighted by the system temperature; and the internal production of entropy. The latter is fundamentally related to the spontaneous evolution ("Verwandlungen"[1]) of the system, and states the universal irreversibility of spontaneous transitions. The relationship of entropy to the second law is often beset with confusion between external transfers of entropy and its internal production.

Clausius[2] defined the extensive variable to which he gave the name "entropy," in terms of its changes:

$$dS = dS_e + dS_i = \frac{dQ}{T} + \frac{dQ'}{T}. \tag{1}$$

Because entropy is defined as an extensive variable, its change with time is determined by two distinct mechanisms: the flow of entropy to or from its defining volume (the thermodynamic system under consideration); and its internal production. Clausius originally called *dQ'* the change in "uncompensated heat;" for he understood that there is always in a thermodynamic transformation a quantity of heat evolved which is intrinsically unrelated to whatever heat that the system may absorb from or transfer to the rest of the universe, *dQ*. Clausius then went on to enunciate the second law as:

$$\boxed{dQ' \geq 0}. \tag{2}$$

In enunciating the second law of thermodynamics, Clausius quite clearly stated that, while the first type of entropy change (that effected by exchange of heat with its surroundings) can be positive, negative or zero, the second type of entropy change (that caused by the internal creation of entropy) *can be only positive* in any spontaneous transformation.

Boltzmann[3] later defined the entropy of an isolated system (for which $dQ = 0$) as an absolute, extensive, thermodynamic variable in terms of the sum of possible configurations, $W$, which the system can assume consistent with its thermodynamic constraints:

$$\boxed{S = k \ln W}. \tag{3}$$

Equation (3) remains the most fundamental and simple definition of entropy, and is engraved on Boltzmann's tombstone. This definition of entropy, (3), presciently acknowledges the absolute quality of entropy. Because $W$ is a natural number, and $\geq 1$, the inequality $S \geq 0$ follows.

The understanding of entropy is sometimes obscured by frequent use of the Boltzmann-Gibbs canonical ensemble, and thermodynamic limit. In a semi-classical approximation,

$$W(E,N,V) = \int_{E}^{E+\varepsilon_0} \frac{d^{3N}\vec{p}\,d^{3N}\vec{q}}{N!(2\pi\hbar)^N} \rho(\vec{q},\vec{p}) \approx \varepsilon_0 \int \frac{d^{3N}\vec{p}\,d^{3N}\vec{q}}{N!(2\pi\hbar)^N} \delta(E - H(\vec{q},\vec{p})), \tag{4}$$

in which $E$ is the total energy, $N$ is the number of particles and $V$ the volume. In equation (4), the density function, $\rho(p,q)$, is that which defines the microcanonical ensemble:

$$\rho(\vec{p},\vec{q}) = \begin{cases} 1 & \text{if} \quad E \leq H(\vec{q},\vec{p}) \leq E + \varepsilon_0 \\ 0 & \text{otherwise} \end{cases}. \tag{5}$$

For a finite quantum-mechanical system:

$$W(E,N,V) = \sum_n \left[\text{all eigenstates, } n, \text{ of } H \text{ of fixed } N \text{ \& } V \text{ with } E \leq E_n \leq (E+\varepsilon_0)\right]. \tag{6}$$

In equations (4) and (6), $\varepsilon_0$ represents the macroscopic energy resolution. In equations (3), (4), and (6), there is no need of the thermodynamic limit, nor of concavity, extensivity and homogeneity. In its semi-classical approximation, equation (4), $W(E,N,V,\cdots,)$ simply measures the area of the sub-manifold of points in the $6N$-dimensional phase-space ($\Gamma$-space) with prescribed energy $E$, particle number $N$, volume $V$, as well as any other required time invariant constraints (here suppressed for simplicity). Because Planck defined entropy in this mathematical form, we shall call such as the Boltzmann-Planck principle.

There are various reviews on the mathematical foundations of statistical mechanics, e.g., the detailed and instructive article by Alfred Wehrl.[4] Wehrl shows how the Boltzmann- Planck formulae, equations (3) and (6), can be generalized to the famous definition of entropy in quantum mechanics by von Neumann:[5]

$$S = -tr[\rho \ln \rho]. \tag{7}$$

using general densities, $\rho$, (which were not necessarily projection-like).

Wehrl discusses the conventional, canonical, Boltzmann-Gibbs statistics where all constraints are fixed only to their mean, allowing for free fluctuations. These free, unrestricted fluctuations of the energy imply an uncontrolled energy exchange with the universe, $dQ$ in Clausius' definition, (1). This assumption, however, is dangerous; for there are situations where the fluctuations are macroscopic and do not vanish in the thermodynamic limit. An example are phase transitions where $\Delta E = E_{\text{latent}}$, the latent heat of transformation. Wehrl points to many serious complications with this definition. However, in the case of conserved variables, more than their mean is known; these quantities are known sharply. In microcanonical thermodynamics, von

Neumann's definition, (7), is not needed; and the analysis can be developed on the level of the original, Boltzmann-Planck definition of entropy, equations (3) and (6). We thus explore statistical mechanics and entropy at their most fundamental level. Because such analysis does not demand scaling or extensivity, it can further be applied to the much wider group of non-extensive systems from nuclei to galaxies[6] and can address the original object for which thermodynamics was enunciated some 150 years ago: phase separations.

The Boltzmann-Planck formula has a simple but deep physical interpretation: $W$ or $S$ are measures of a lack of precision specifying the complete set of initial values for all $6N$ microscopic degrees of freedom needed to specify unambiguously the $N$-body system.[7] Usually only a few time-independent control parameters $E$, $N$, $V$,···, of the system are known, which are conserved or very slowly varying; and there exists no control of the other fast changing degrees of freedom. If specification of the system were complete, - i.e., if all $6N$-degrees of freedom sharply known at some time $t_0$, - $W$ would be a single cell of size $(2\pi\hbar)^{3N}$ in the $6N$-dimensional phase space, and $S$ would be 0.

Many initial values for all $6N$ necessary microscopic variables are consistent with the same values of $E,N,V$,···; thus the system has intrinsically a redundancy of control parameters. All other degrees of freedom vary with time, usually rapidly, and are normally not under control. The manifold of all these points in the $6N$-dimensional phase space is the microcanonical ensemble, which has a well-defined geometrical size $W$ and, by equation (3), a nonvanishing entropy, $S(E,N,V,\cdots)$. The dependence of $S(E,N,V,\cdots)$ on its arguments determines completely thermostatics and equilibrium thermodynamics.

Clearly, Hamiltonian (Liouvillean) dynamics of the system cannot create the missing information about the initial values, - i.e., the entropy, $S(E,N,V,\cdots)$, cannot decrease. As has been worked out,[6,8] the inherent finite resolution of the macroscopic description implies an increase of $W$ or $S$ with time when an external constraint is relaxed. Such constitutes a statement of the second law of thermodynamics, which requires that the internal production of entropy be positive for every spontaneous process. Analysis of the consequences of the second law by the microcanonical ensemble is appropriate because, in an isolated system (which is the one relevant for the microcanonical ensemble), the changes in total entropy must represent the internal production of entropy, and there are no additional, uncontrolled, fluctuating, energy exchanges with the environment.

It should be emphasized that this proper statistical definition of the entropy, $S(E,N,V,\cdots)$ characterizes the whole microcanonical ensemble,[7] for it measures the total number of the possible states of the system under the information given. Equations, (3)-(6) are in clear contrast to the definition suggested by:[9-11]

$$W_{\text{bulk}}(E,N,V) = \int \frac{d^{3N}\vec{p}\,d^{3N}\vec{q}}{N!(2\pi\hbar)^N} \Theta(E - H(\vec{q},\vec{p})) = \exp(S_{\text{bulk}}(E,N,V))$$
$$= \sum_n [\text{all eigenstates, } n, \text{ of } H \text{ with } E_N \leq E]$$
(8)

This statistical definition, (4) and (6) (called $S_{\text{surf}}$ in[11]), has *a priori* not much to do with the adiabatic invariants of a single trajectory in the *6N*-dimension. phase space.

As will be shown in section 3, the entropy, $S(E)$, can be convex or concave with respect to $E$ for non-extensive systems, - in contrast to the assumption made by Hertz.[10] In such cases, the temperature, or the mean kinetic energy per particle, does not control the energy flow during the equilibration of two systems in thermal contact, - as was the crucial argument of Hertz to introduce $S_{\text{bulk}}$, (8). Contrastingly, the Boltzmann entropy, (4), still controls the direction of the energy flow under equilibration even in these somewhat counter-intuitive, though ubiquitous, situations. This clearly emphasizes that the Boltzmann entropy, (3), is the fundamental quantity for statistical mechanics, - *not* the temperature. (This statement refutes contradictory assertions made previously.[9-11])

Moreover, the definition of the bulk entropy $S_{bulk}(E,N,V)$, (8), contains also energetically inaccessible states, characterized by less energy, which are clearly excluded by our knowledge of the total energy. Thus $S_{bulk}(E,N,V)$ does not measure the redundancy, or ignorance, of the available information, and is therefore unacceptable for conceptual reasons. This is a clear warning that one should not follow some formal definition and disregard the original clear meaning of entropy and statistics, as is too often done in the literature. Of course, for homogeneous macroscopic systems, the difference between the two alternative definitions of the entropy equations (4) and (6), and equation (8) disappears.

## 2. The 0$^{th}$ Law in conventional, extensive thermodynamics.

This section and the following discuss mainly systems that have no macroscopic (extensive) control parameters except energy; the particle density does not change, and there are no chemical reactions. In conventional (extensive) thermodynamics, the thermal equilibrium of two systems is established by bringing them into thermal contact and allowing exchange of free energy. Equilibrium is established when the total entropy,

$$S_{total}(E,E_1) = S_1(E_1) + S_2(E - E_1) \qquad (9)$$

is maximal, such that

$$dS_{total}(E,E_1)|_E = dS_1(E_1) + dS_2(E - E_1) = 0. \qquad (10)$$

Under an energy flux $\Delta E_{2\to 1}$ from 2→1, the total entropy changes to lowest order in $\Delta E$ by

$$\Delta S_{total} = (\beta_1 - \beta_2)\Delta E_{2\to 1}, \qquad (11)$$

in which

$$\beta = \frac{dS}{dE} = \frac{1}{T}. \tag{12}$$

Consequently, a maximum of $S_{total}(E = E_1 + E_2, E_1)|_E$ will be approached when

$$sign(\Delta S_{total}) = sign(T_2 - T_1) sign(\Delta E_{2\to 1}) > 0, \tag{13}$$

From here, Clausius' first formulation of the Second Law follows: "Heat always flows spontaneously from hot to cold." Essential for this conclusion is the additivity of $S$ under the separation of the two subsystems, equation (9). Thus, temperature is an appropriate control parameter for extensive systems.

### 3. Stability against spontaneous energy gradients: There exists no phase separation without a convex, non-extensive entropy, $S(E)$.

Small systems and very large self-gravitating systems are inhomogeneous,[12] as are essentially also systems at phase separation. For these cases, the additivity and extensivity of $S$, and also of $E$, do not hold.[12] The main purpose to develop this new and extended version of thermodynamics[12] is to address such systems. The Boltzmann-Planck entropy, equation (6), does not scale linearly with the volume or number of particles, but has an important, *non-linear*, correction which we shall call henceforth $\Delta S_{\text{surf-corr}}$.

Since the beginning of thermodynamics in the 19th century, its principle motivation was to describe steam engines and the gas-liquid phase transition of water. In the gas-liquid phase transition, water becomes inhomogeneous and develops a separation of the gas phase from the liquid, - i.e. it boils. As conventional canonical statistical mechanics works only for *homogeneous, infinite* systems, phase separations remain outside standard Boltzmann-Gibbs statistical thermodynamics and are identified by Yang-Lee singularities.

At phase separation in general, the microcanonical caloric curve *T(E)* is *backbending*. Here the heat capacity, $C = (\partial T/\partial E)^{-1} = - (\partial S/\partial E)^2/(\partial^2 S/\partial E^2)$ becomes negative, the curvature $\partial^2 S/\partial E^2$ is positive, and *S(E)* is convex.

The weight, $e^{(S(E)-E/T)}$, of the configurations with energy *E* in the definition of the canonical partition sum,

$$Z(T) = \int_0^\infty e^{(S(E)-E/T)} dE, \tag{14}$$

becomes bimodal; at the transition temperature, it has two peaks, the liquid and the gas configurations which are separated by the latent heat. Consequently *S(E)* must be convex and the weight in (14) has a minimum between the two pure phases. Of course, the minimum can only be seen in the microcanonical ensemble where the energy is controlled and its fluctuations forbidden. Otherwise, in a canonical ensemble, the system would fluctuate between the two pure phases separated by an energy difference, $\Delta E \approx E_{latent}$, on the order of the latent heat. These fluctuations thus scale proportionately to *N* (not proportionately to √*N*) and do not vanish in the thermodynamic limit. Thus, the convexity of *S(E)* is the generic signal of a phase transition of first order and of phase-separation.[12] Anomalously, this fact, essential for the original purpose of thermodynamics, which was to describe steam engines, has never been treated completely during the past 150 years.

The ferromagnetic Potts-model illuminates in a most simple example the occurrence of a back-bending caloric curve *T(E)*.[13] A typical plot of $s(e,N) = S(E=Ne)/N$ in the region of phase separation is shown in fig(1). Consider the two equal systems depicted in Fig. 1 to be combined, such that one has the energy per lattice site $e_a = e_1 - \Delta e$ at temperature $T_a$, and the other has energy per lattice site $e_b = e_3 + \Delta e$ at temperature $T_b$; where $T_b < T_a$. Consider also that energy is allowed to be exchanged without constraint. Then the systems will equilibrate at energy $e_2$ with an increase of entropy. The temperature $T_a$ drops(cooling) and energy, in the form of heat, flows (on the average) from $b \to a$. Thus heat flows spontaneously from the colder system to the hotter one; and the usual Clausius formulation of the Second Law is violated. Although this property is well known for self-gravitating systems, it is not solely a peculiarity of such. This situation is generic at phase separations within classical thermodynamics, even for systems with short-range coupling and has no intrinsic connection with long range interactions. In Section IV the general microscopic reasons for this convexity are discussed.

An important point to be noted is that the back-bending of $T(e)$ does *not* depend upon the periodic boundary condition used here. Numerous realistic microcanonical calculations for nuclear fragmentation and atomic-cluster fragmentation[12,14] have shown such back-bending without use of periodic boundary conditions. These results have

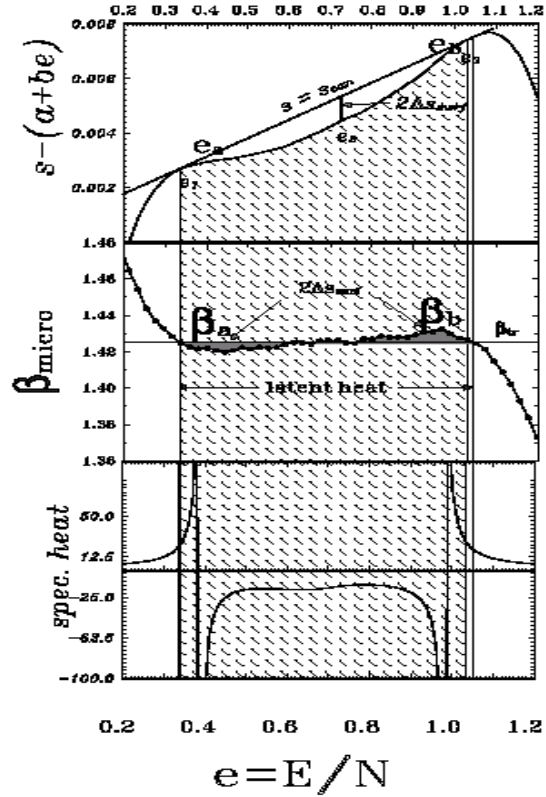

**Fig. 1 Ferromagnetic Potts model ($q = 10$) on a 50×50-lattice with periodic boundary conditions in the region of phase separation. At the energy $e_1$ per lattice point, the system is in the pure ordered phase, at $e_3$ in the pure disordered phase. At $e_a$, slightly above $e_1$, the temperature $T_a = 1/\beta_a$ is higher than $T_2$ and more so than $T_b$ at $e_b$, slightly below $e_3$. At $e_a$, the system separates into a few bubbles of the disordered phase embedded in the ordered phase, at $e_b$ into a few droplets of the ordered phase within the disordered one.**

been verified by numerous experimental data in nuclear[15] and cluster physics.[16] The macroscopic energy fluctuations and the resulting negative specific heat in high-energy physics were described early by Carlitz.[17] (Previous errors connected with the use of periodic boundary conditions[18] have been corrected.[19])

Consider the system separated into two pieces, a & b, by a dividing surface, with half the number of particles each. The dividing surface is purely geometric; it exists only as long as the two pieces can be distinguished by their different energy/particle $e_a$ and $e_b$. If the constraint on the difference $e_b$-$e_a$ is fully relaxed, and $e_b$-$e_a$ can fluctuate freely at fixed $e_2 = (e_a + e_b)/2$, the dividing surface may be assumed to have no further physical effect on the system. Constraining the energy-difference ($e_a$ - $e_b$) between the two, reduces the number of free, unconstrained degrees of freedom and reduces the entropy by -2$\Delta S_{surf-corr}$. By constraining the system in such fashion, one adds information and reduces the entropy. (If the effect of the surface were also to cut some chemical bonds, the entropy would be reduced by the separation. Before the separation, there were configurations with attractive interactions across the surface which are interrupted by the division; their energy level would shift upwards outside the permitted band-width $\varepsilon_0$, and be eliminated from the partition sum (6).)

For an *extensive* system $S(E,N) = Ns(e = E/N) = S(E/2,N/2)$. One can argue as follows: The combination of two pieces of $N/2$ particles each, one at $e_a = e_2 - \Delta e/2$ and a second at $e_b = e_2 + \Delta e/2$ must lead to $S(E_2,N) \geq [S(E_a/2,N/2) + S(E_b/2,N/2)]$, the simple algebraic sum of the individual entropies, because by combining the two pieces one normally looses information. This, however, is equal to $[S(E_a,N) + S(E_b,N)]/2$ thus $S(E_2,N) \geq [S(E_a,N)+S(E_b,N)]/2$, i.e., *the entropy S(E,N) of an extensive system is necessarily concave.*

For a *non-extensive* system, one has in general $S(E,N) \geq 2S(E/2,N/2)$, because again two separated, closed pieces have more information than their unification. Now, if $E_2$ is the point of maximum positive curvature of $S(E,N)$ (convexity = upwards-concave as $y = x^2$), one has $S(E_2,N)$ $\leq [S(E_a,N) + S(E_b,N)]/2$ as in Fig. 1. However, the right side is larger than $[S(E_a/2,N/2) + S(E_b/2,N/2)]$; i.e. even though $S(E,N)$ is convex at constant $N$, the unification of the pieces with $S(E_a/2,N/2)$ and $S(E_b/2,N/2)$ can still lead to a *larger* entropy $S(E_2,N)$. The difference between $[S(E_a,N) + S(E_b,N)]/2$ and $[S(E_a/2,N/2) + S(E_b/2,N/2)]$ we shall call henceforth $\Delta S_{surf-corr}$. The correct entropy balance, before and after establishing the energetic split $e_b > e_a$ of the system, is

$$S_{\text{after}} - S_{\text{before}} = \frac{S_a + S_b}{2} - \Delta S_{\text{surf-corr}} - S_2 \leq 0, \tag{15}$$

even though the difference of the first and the last term is positive.

Consider a transformation in the inverse direction: By *relaxing* the constraint and allowing, on average, for an energy flux, $\Delta E_{b \to a} > 0$, *against* $T_a - T_b > 0$, (i.e., against the temperature gradient but in the direction of the decreasing energy gradient), an increase of $S_{\text{total}} \to S_2$ obtains. This result is consistent with the normal picture of energy equilibration. Thus the Clausius statement, "*energy flows always from hot to cold*," which is a statement of the dominant role of the temperature in thermo-statistics,[10] is violated.

These considerations show that, unlike extensive thermodynamics, the temperature is not an appropriate control parameter in non-extensive systems. In the thermodynamic limit of a system with short-range coupling, $N \to \infty$ as $\Delta S_{\text{surf-corr}} \sim N^{2/3}$, and therefore $(\Delta S_{\text{surf-corr}})/N \sim N^{-1/3}$, must go to 0 in accordance with van Hove's theorem.

## 4. The microcanonical mechanism leading to condensation, phase separation, and the origin of the convexities of S(E).

Many applications of microcanonical thermodynamics to realistic examples of hot nuclei, atomic clusters, and rotating astrophysical systems have been presented during the past twenty years which demonstrate convex intruders in the microcanonical entropy and, consequently, negative heat capacities. Such are reviewed in the publication list on the web site http://www.hmi.de/people/gross/ and elsewhere.[20]

Here following are illuminated the general microscopic mechanism leading to the appearance of a convex intruder in $S(E,V,N,\cdots)$. This is the generic signal of phase transitions of first order and of phase-separation within the microcanonical ensemble.

Assume the system is classical and obeys the following Hamiltonian:

$$\left. \begin{array}{l} H = \sum_i^N \dfrac{p_i^2}{2m} + \Phi^{int}(\{\vec{r}\}) \\ \Phi^{int}(\{\vec{r}\}) = \sum_{i>j=1}^N \phi(\vec{r}_i - \vec{r}_j) \end{array} \right\} \quad (16)$$

In this case the system is controlled by energy and volume.

### 4.1. The gas-liquid phase transition.

The microcanonical sum of states or partition sum is:

$$\begin{aligned} W(E,N,V) &= \dfrac{1}{N!(2\pi\hbar)^{3N}} \int_{V^N} d^{3N}\vec{r} \int d^{3N}\vec{p}\, \varepsilon_0 \delta\!\left(E - \sum_i^N \dfrac{\vec{p}_i^2}{2m_i} - \Phi^{int}(\{\vec{r}\})\right) \\ &= \dfrac{V^N \varepsilon_0 (E-E_0)^{(3N-2)/2} \prod_1^N m_i^{3/2}}{N!\,\Gamma(3N/2)(2\pi\hbar^2)^{3N/2}} \int_{V^N} \dfrac{d^{3N}\vec{r}}{V^N}\!\left(\dfrac{E - \Phi^{int}(\{\vec{r}\})}{(E-E_0)}\right)^{(3N-2)/2}; \\ &= W^{IG}(E-E_0,N,V) \times W^{int}(E-E_0,N,V) \\ &= \exp\!\left[S^{IG} + S^{int}\right] \end{aligned} \quad (17)$$

in which

$$W^{\text{IG}}(E-E_0,N,V) = \frac{V^N \varepsilon_0 (E-E_0)^{(3N-2)/2} \prod_{1}^{N} m_i^{3/2}}{N!\Gamma(3N/2)(2\pi\hbar^2)^{3N/2}}, \qquad (18)$$

and

$$W^{\text{int}}(E-E_0,N,V) = \int_{V^N} \frac{d^{3N}\vec{r}}{V^N} \Theta\big(E-\Phi^{\text{int}}(\{\vec{r}\})\big) \left(1 - \frac{\Phi^{\text{int}}(\{\vec{r}\})-E_0}{(E-E_0)}\right)^{(3N-2)/2}. \qquad (19)$$

$V$ is the spatial volume; $E_0 = \min(\Phi^{\text{int}})$ is the energy of the ground-state of the system.

The separation of $W(E,N,V)$ into $W^{\text{IG}}$ and $W^{\text{int}}$ is the microcanonical analogue of the split of the canonical partition sum into a kinetic part and a configuration part:

$$Z(T) = \frac{V^N}{N!}\left(\frac{mT}{2\pi\hbar^2}\right)^{3N/2} \int_{V^N} \frac{d^{3N}\vec{r}}{V^N} \exp\left(\frac{-\Phi^{\text{int}}(\{\vec{r}\})}{T}\right). \qquad (20)$$

In the thermodynamic limit, the order parameter of the (homogeneous) liquid-gas transition is the density. The transition is linked to a condensation of the system towards a larger density controlled by pressure. For a finite system, we expect analogous behavior. However, for a finite system, the transition is controlled by the constant system volume $V$. At low energies, the $N$ particles condensate into a droplet with much smaller volume $V_{0,N}$. $3(N-1)$ internal coordinates are limited to $V_{0,N}$. Only the center of mass of the droplet can move freely in $V$ (remember we did not fix the center-of-mass in equation (17)). The system does not fill the $3N$-configuration space $V^N$. Only a stripe with width $(V_{0N})^{1/3}$ in $3(N-1)$ dimensions of the total $3N$-dimension space is populated. The system is *non-homogeneous* even though it is equilibrated and, at low energies, internally in the single liquid phase; and it is not characterized by an intensive homogeneous density. In fact, $W^{\text{int}}(E-E_0,N,V)$ can be written as:

$$W^{\text{int}}(E-E_0,N,V) = \left(\frac{V(E,N)}{V}\right)^N \leq 1. \qquad (21)$$

$$\left(V(E,N)\right)^{N} \equiv \int_{V^{N}} d^{3N}\vec{r}\,\Theta\!\left(E-\Phi^{\text{int}}(\{\vec{r}\})\right)\left(1-\frac{\Phi^{\text{int}}(\{\vec{r}\})-E_{0}}{(E-E_{0})}\right)^{(3N-2)/2}. \tag{22}$$

$$S^{\text{int}}(E-E_{0},N,V) = N\ln\!\left(\frac{V(E,N)}{V}\right) \leq 0. \tag{23}$$

The first factor, $\Theta\!\left(E-\Phi^{\text{int}}(\{r\})\right)$, in equation (22) sorts the energetically forbidden region from the potential holes (clusters) in the 3N-dimension potential surface $\Phi^{\text{int}}(\{\vec{r}\})$. The volume $V^{N}(E,N) \leq V^{N}$ is the accessible part of the 3N-dimension-spatial volume (potential holes) outside of the forbidden regions which have $\Phi^{\text{int}}(\{\vec{r}\}) > E$. Thus $V^{N}(E,N)$ is total the 3N-dimensional eigen-volume of the condensate (droplets), with $N$ particles at the given energy, summed over all possible partitions, clustering, in 3N-configuration space. The relative volume of each partition compared with $V^{N}(E,N)$ gives its relative probability. $V^{N}(E,N)$ has the limiting values:

$$\left[V(E,N)\right]^{N} = \begin{cases} V^{N} & \text{for } E \text{ in the gas phase,} \\ V_{0,N}^{N-1}V & \text{for } E = E_{0}. \end{cases} \tag{24}$$

$W^{\text{int}}(E-E_{0},N,V)$ and $S^{\text{int}}(E-E_{0},N,V)$ have the limiting values:

$$W^{\text{int}}(E-E_{0}) \leq 1, \;\Rightarrow\; S^{\text{int}}(E-E_{0},N) \leq 0$$

$$\to \begin{cases} 1 & \text{for } E \gg \Phi^{\text{int}}, \\ \left(\dfrac{V_{0N}}{V}\right)^{N-1} & \text{for } E \to E_{0}. \end{cases} \tag{25}$$

$$S^{\text{int}}(E-E_{0}) \to \begin{cases} 0 & \text{for } E \gg \Phi^{\text{int}}, \\ \ln\!\left(\dfrac{V_{0N}}{V}\right)^{N-1} < 0 & \text{for } E \to E_{0}. \end{cases} \tag{26}$$

All physical details are contained in $W^{\text{int}}(E-E_{0},N,V)$ or $S^{\text{int}}(E-E_{0},N,V)$, alias $N\ln[V(E,N)]$, c.f. eqs.(21)-(26): If the energy is high the detailed structure of $\Phi^{\text{int}}(\{\vec{r}\})$ is unimportant $W^{\text{int}}$ $\approx 1$, $S^{int} \approx 0$. The system behaves like an ideal gas and fills the volume $V$. At sufficiently low energies only the minimum of $\Phi^{int}(\{\vec{r}\})$ is explored by $W^{int}(E-E_0,N,V)$. The system is in a condensed phase, a single liquid drop, which moves freely inside the empty larger volume $V$, the internal degrees of freedom are trapped inside the reduced volume $V_{0N} \ll V$.

One can guess the general form of $N\ln[V(E,N)]$: Near the ground state $E >\sim E_0$, it must be flat $\approx (N-1)\ln(V_{0N}) + \ln(V)$ because the liquid drop has some eigen-volume $V_{0N}$ in which each particle can move (liquid). With rising energy, $\ln[V(E,N)]$ increases to the point ($E^{trans}$), where it is first possible that a drop fissions into two. Here an additional new configuration opens in the $3N$-dimension configuration space: Either one particles evaporates from the cluster and explores the larger external volume $V$, or the droplet fissions into two droplets and their respective two center-of-mass coordinates explore the larger $V$. This gives a sudden upward jump in $S^{int}(E)$ by an amount on the order of $\ln[(V - V_{0(N-1)})/V_{0(N-1)}]$ and similar in the second case.

Of course, this sudden opening of additional parts of the $3N$-dimension configuration space gives also a similar jump upwards in the total entropy:

$$S(E) = S^{IG} + S^{int}$$
$$\propto \ln\left[\int_{V^N} d^{3N}\vec{r}\, \Theta\left(E - \Phi^{int}(\{\vec{r}\})\right)\left(E - \Phi^{int}(\{\vec{r}\})\right)^{(3N-2)/2}\right]. \tag{27}$$

by $\ln[(V - V_{0(N-1)})/V_{0(N-1)}]$. Thus, the magnitude of the increase of $S(E)$ is controlled by the system volume, $V$. Later further such "upwards jumps" may follow. Each of these "jumps" induce a convex upwards bending of the total entropy $S(E)$, equation (27). Each is connected to a bifurcation and bimodality of $\exp[S(E) - E/T]$ and the phenomenon of phase-separation. In the conventional canonical picture for a large number of particles this is forbidden and hidden behind the familiar Yang-Lee singularity of the liquid to gas phase transition.

In the microcanonical ensemble this is analogue to the phenomenon of multi-fragmentation in nuclear systems.[12,21] This, in contrast to the mathematical Yang-Lee theorem, physical microscopic explanation of the liquid to gas phase transition sheds sharp light on the physical origin of the transition, the sudden change in the inhomogeneous population of the $3N$-dimensional configuration space.

### 4.2. The liquid-solid phase transition.

In contrast to the liquid phase, in the crystal phase a molecule can only move locally within its lattice cage of the size $d^3$ instead of the whole volume $V_{0N}$ of the condensate. I.e. in equation (26) instead we have $S^{int} \to \ln[(d^3/V_{0,N})^{N-1}]$.

### 4.3. Summary of section IV.

The gas-liquid transition is linked to the transition from uniform filling of the container volume, $V$, by the gas to the smaller eigen-volume of the system, $V_0$, in its condensed phase when the system is *inhomogeneous* (some liquid drops inside the larger empty volume $V$). First $3(N-1)$, later at higher energies less and less degrees of freedom condensate into a drop. First three, then more and more degrees of freedom (center-of-mass-coordinates of the drops) explore the larger container volume $V$ leading to incremental increases (convexities) in $S^{int}(E)$. The volume of the container controls how close the system is to the critical end-point of the transition, where phase-separation disappears. Towards the critical end-point, i.e. with smaller $V$, the jumps $\ln[V]-\ln[V_0]$ become smaller and smaller. In the case of the solid-liquid transition, however, the external volume, $V$, of the container confines only the center-of-mass position of the crystal, respectively, the droplet. The entropy jumps during melting by $\Delta S^{int} \propto \ln[V_{0,N}/d^3]$.

page 18 of 23

At the surface of a drop $\phi^{int} > E_0 = \min[\phi^{int}]$, i.e. the surface gives a negative contribution to $S^{int}$ in equation (22) and to $S$ at energies $E \geq E_0$, as was similarly assumed in section 3 and explicitly in equation (15).

**Acknowledgement.**

D. H. E. Gross is grateful to J. Möller for insistent, therefore helpful, discussions.

**Appendix A. A simple model for condensation.**

Assume the various potential-pockets are attractive square-wells like with depths $\Phi_\lambda < 0$. With the (somewhat schematic) abbreviation:

$$I(\mathbf{K}) := \int_{V^N} \frac{d^{3N}\vec{r}}{V^N} \Theta\left(E - \Phi^{int}(\{\vec{r}\})\right) \left(E - \Phi^{int}(\{\vec{r}\})\right)^{(3N-\mathbf{K})/2}$$
$$= \sum_{\lambda=1}^{N} \left\langle (E - \Phi_\lambda)^{(3N-\mathbf{K})/2} \right\rangle \left(V - \sum_{k=1}^{N-\lambda} V_{0k}\right)^{\lambda} \prod_{k=1}^{N-\lambda} V_{0k} \qquad (28)$$

Here $\Phi_\lambda \leq 0$ are different topological regions (potential pockets) of $\Phi^{int}(\{r\})$: Single cluster ($\lambda = 1$), two clusters ($\lambda = 2$), several clusters or, finally, if energetically possible, ($\lambda = N$) free particles, $\Phi = 0$. Depending on $\lambda$, there are $3\lambda$ center-of-mass coordinates which can move over the whole volume $V$ in contrast to the rest $3(N - \lambda)$ which are limited to smaller cluster volumes $V_{0k}$.

All particles condensed into single clusters $\lambda = 1$ will have the lowest potential $\Phi_1$ and consequently the largest $E - \Phi$. On the other hand the volume factor $\sim V/V_{0(N-1)}$ appears only once. Then other terms with $\lambda > 1$ may dominate if $(V/V_{0(N-\lambda)})^\lambda \gg (V/V_{0(N-1)})$.

In terms of $I(\mathbf{K})$, the partition function $W(E)$ and its derivatives are expressed as:

$$W(E, N, V) \propto I(2). \qquad (29)$$

$$\frac{\partial W(E,N,V)}{\partial E} \propto \frac{3N-2}{2} I(4). \tag{30}$$

$$\frac{\partial^2 W(E,N,V)}{\partial E^2} \propto \frac{(3N-2)(3N-4)}{4} I(6). \tag{31}$$

$$\beta(E) \propto \frac{3N-2}{2} \frac{I(4)}{I(2)}. \tag{32}$$

$$\beta'(E) \propto \frac{3N-2}{2} \left( \frac{3N-4}{2} \frac{I(6)}{I(2)} - \frac{3N-2}{2} \left( \frac{I(4)}{I(2)} \right)^2 \right). \tag{33}$$

To get a convexity ($\beta' \geq 0$), we must have:

$$\frac{I(6)I(2)}{|I(4)|^2} > 1 + \frac{2}{3N-4} \tag{34}$$

**Appendix B. Some examples**

**B.1.** **A single phase.**

When calculating equations (32)-(34), one immediately sees that for a sharply mono-dispersed sum (28), $\lambda \sim \lambda_0$ (single phase):

$$I(\mathbf{K}-2) \sim (E - \Phi_{\lambda_0}) I(\mathbf{K}) \tag{35}$$

Then left side of criterion (34) becomes 1 and there is no convexity.

$$\beta(E) = \frac{3N-2}{2} \frac{1}{E - \Phi_{\lambda_0}}. \tag{36}$$

$$\beta'(E) = -\frac{2}{3N-2} \beta^2. \tag{37}$$

I.e. the heat capacity is:

$$Nc(E) = -\frac{\beta^2}{\beta'} = \frac{3N-2}{2}. \tag{38}$$

equal to the ideal gas value, independently of whether the system is condensed or not. (Remember, in equation (28) we assumed for simplicity that $\Phi_\lambda$ has a flat bottom).

### B.2. The bimodal distribution, phase-separation.

When the mass dispersion (28) is bimodal when e.g. two terms in the sum dominate, i.e. when

$$I(\mathbf{K}) \approx \sum_{\lambda=\lambda_1,\lambda_1} \left\langle (E-\Phi_\lambda)^{(3N-\mathbf{K})/2} \right\rangle V^\lambda \prod_{k=1}^{N-\lambda} V_{0k} . \tag{39}$$

then the different I(K) are not anymore proportional for different *K* as in equation (35) and equation (33) can very well be positive. As anticipated in our previous discussion in section (IV), the volume-factor, $V/V_0 > 1$, determines whether the system can perform a phase-transition and whether the transition has a critical end-point and if, where it is, i.e. where the region of convexity disappears.

### B.3. Hard disks, the Alder-Wainwright Transition.[22]

For hard disks the potential is:

$$d = |\mathbf{r}-\mathbf{r}'| \tag{40}$$

$$\phi(d) = \begin{cases} \infty & d \leq d_{disk} \\ 0 & d > d_{disk} \end{cases} \tag{41}$$

$$\Phi^{int}(\{\mathbf{r}\}) = \sum_{i>j} \phi(\vec{r}_i - \vec{r}_j) \tag{42}$$

$$E_0 = 0 \tag{43}$$

$$\mathbf{I}(K) = E^{(3N-K)/2} \sum_\lambda (V-V_{0,N})^\lambda V_{0,N}^{N-\lambda} \tag{44}$$

Due to the criterion (34) there is no bimodality vs. energy, i.e. no latent heat, and the heat capacity is everywhere the ideal-gas value $N_c(E) = (3N-2)/2$.

The only interesting change of $W$ is as function of the external volume $V$. To get some feeling for this, we make the following simplified ansatz: We assume each cluster with $n$ disks has the close-packing volume $V_n = (n/N)V_0$. Moreover, each cluster has a *constant* surface layer of volume $d$ around to separate it from other clusters. In a configuration of $\lambda$ clusters there are $3\lambda$ positional (cluster) degrees of freedom which can move free in volume of $(V-V_0-\lambda d)^\lambda$ whereas $3(N-\lambda)$ internal degrees of freedom are limited to the condensate volume $V_0^{(N-\lambda)}$.

For number of states we have after these simplifying assumptions, cf. equation (44):

$$\left.\begin{array}{l} W \propto \sum_\lambda W_\lambda \\ W_\lambda = (V-V_0-\lambda d)^\lambda (V_0)^{N-\lambda} \end{array}\right\} \quad (45)$$

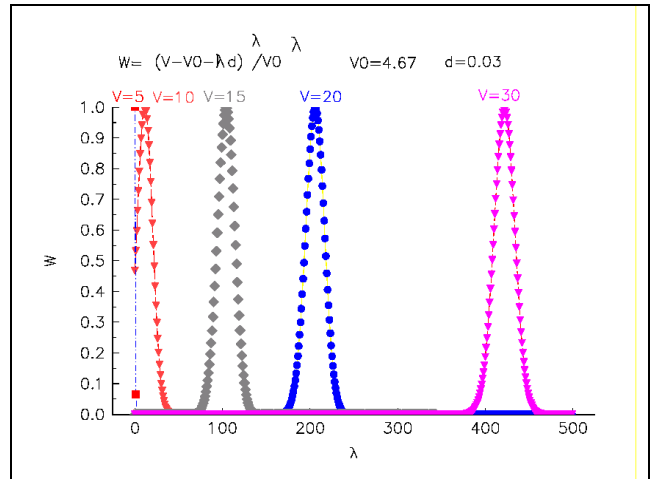

The form of $W_\lambda$ as function of the external volume is shown in fig(2): Evidently, $W_\lambda(V)$ is for every $V$ a sharp mono-dispersed function of number of clusters and shows no bimodality vs. $\lambda$. Consequently there is no phase separation in $P(V)$ curve. This is not a phase-transition of first order.

**Fig. 2 Simplified Alder-Wainwright transition, with normalized cluster-number, $\lambda$, distributions as functions $W_\lambda(V)$ of external volume, $V$.**